\documentclass[runningheads]{svmult}

\usepackage{makeidx}   
\usepackage{graphicx}  
\usepackage{subeqnar}  
\usepackage{multicol}  
\usepackage{physprbb}  
\makeindex             

\newcommand{\ttheta}{\mbox{\boldmath $\theta$} {}}
\newcommand{\EQ}{\begin{equation}}
\newcommand{\EN}{\end{equation}}
\newcommand{\EQA}{\begin{eqnarray}}
\newcommand{\ENA}{\end{eqnarray}}

\newcommand{\Eqs}[2]{eqs.~(\ref{#1}) and~(\ref{#2})}

\newcommand{\Fig}[1]{figure~\ref{#1}}

\newcommand{\bra}[1]{\langle #1\rangle}

\newcommand{\uu}{\mbox{\boldmath $u$} {}}

\newcommand{\BB}{\mbox{\boldmath $B$} {}}

\newcommand{\AAA}{\mbox{\boldmath $A$} {}}

\newcommand{\JJ}{\mbox{\boldmath $J$} {}}

\newcommand{\ff}{\mbox{\boldmath $f$} {}}

\newcommand{\SSS}{\mbox{\boldmath $S$} {}}

\newcommand{\nab}{\mbox{\boldmath $\nabla$} {}}

\newcommand{\oo}{\mbox{\boldmath $\omega$} {}}

\newcommand{\grad}{{\rm grad} \, {}}

\newcommand{\dd}{{\rm d} {}}

\def\half{{\textstyle{1\over2}}}

\newcommand{\K}{\,{\rm K}}

\begin{document}
\title*{Helicity in Hydro and MHD Reconnection}
\toctitle{Helicity in MHD and Hydro Reconnection}
\titlerunning{Helicity and Reconnection}
\author{Axel Brandenburg\inst{1,2}
\and Robert M. Kerr\inst{3,4}}
\authorrunning{A. Brandenburg and R. M. Kerr}
\institute{Department of Mathematics, University of Newcastle upon Tyne,
NE1 7RU, UK
\and NORDITA, Blegdamsvej 17, DK-2100 Copenhagen \O, Denmark
\and NCAR, Boulder, CO 80307-3000, USA\
\and Department of Atmospheric Sciences, University of Arizona,
Tucson, AZ 85721-0081, USA}

\maketitle              

\begin{abstract}
Helicity, a measure of the linkage of flux lines, has subtle
and largely unknown effects upon dynamics.  Both
magnetic and hydrodynamic helicity are conserved for ideal
systems and could suppress nonlinear dynamics.
What actually happens is not clear because
in a fully three-dimensional system there are additional channels
whereby intense, small-scale dynamics can occur.  This contribution
shows one magnetic and one hydrodynamic case where for each the presence
of helicity does not suppress small-scale intense dynamics of the
type that might lead to reconnection.
\end{abstract}

\section{Introduction}

The term reconnection is used in both the MHD and fluids
communities to describe topological changes in magnetic or vorticity
fields due to resistivity or viscosity and could not occur in
ideal cases where these dissipative terms are zero.
In the strictly ideal limit the
connectivity of the field lines would not change, but this is a singular
limit and even the smallest amount of resistivity or
viscosity allows the connectivity
to change, albeit on small length scales. In the presence of finite
dissipative terms, large amounts of energy can be converted into heat
(or other forms of energy if one goes beyond the hydrodynamic 
approximations). MHD reconnection
plays a role in understanding why the solar corona (i.e.\ the tenuous
layers above the solar surface) are heated to $\sim10^6\K$, even though
at the surface of the sun the temperature is only $\sim6000\K$.
The other aspect is that the dissipative terms allow the field line
connectivity to change. There are strong indications from observations
of the solar corona in X-rays that field loops originally tied to the
solar surface all of a sudden break loose and transport large amounts of
flux into outer space. This raises the issue of how fast can reconnection
occur. This is perhaps the single most challenging aspect of the problem.

Early work on MHD reconnection was concerned with steady state
configurations, allowing a constant flux of material to pass through
an X-point type configuration in two-dimensional field line
configurations. However, because the reconnection site becomes very thin
as the magnetic resistivity decreases, the amount of flux processed
through the reconnection site decreases like the square root of the
resistivity and by this mechanism
finite reconnection in a dynamical timescale is not feasible for typical
astrophysical values of resistivity.
Other more complicated initial conditions can lead to
slow shocks that increase the reconnection rate, as
discussed in a recent textbook \cite{PriestForbes00}.
Is not clear, however, whether the various boundary conditions
studied so far represent anything
physical in the corona and whether the results could explain the nanosecond
timescales over which hard X-ray output associated with reconnection
is seen to rise.

\section{Dissipation of energy and helicity}

Magnetic reconnection has two distinct aspects. One is the speed
at which magnetic energy can be converted into heat and the other
is the speed at which the magnetic topology can change. The two
need not be the same. The perhaps worst possible type of topology
to change is one that invokes mutual linkage of flux tubes, which
can be described by the magnetic helicity $H$ defined as
\EQ
H=\int\AAA\cdot\BB\,\dd V,
\label{eq:hel}
\EN
where $\BB$ is the magnetic field and $\AAA$ is the vector potential such
that $\BB=\nab\times\AAA$. Obviously, $\AAA$ is not uniquely defined,
because adding an arbitrary gradient field to $\AAA$ would not change
$\BB$. However, the value of $H$ is unaffected by this if the integral
is taken over a domain where the normal component of the field vanishes
on the boundaries. In that case
\EQ
\int(\AAA+\grad\varphi)\cdot\BB\,\dd V
=\int\AAA\cdot\BB\,\dd V+\int\varphi\nab\cdot\BB\,\dd V=H,
\EN
because the magnetic field is always solenoidal, $\nab\cdot\BB=0$.
Another conserved quantity is the cross helicity,
$H_{\rm c}=\int\uu\cdot\BB\,\dd V$, which describes the linkage between
flux tubes and vortex tubes. In the absence of magnetic fields
the hydrodynamic helicity, $H_{\rm h}=\int\uu\cdot\oo\,\dd V$,
is conserved by the inviscid Euler equations, and it describes the
linkage of vortex tubes with themselves.

The standard example that highlights the connection between magnetic
helicity and topology is an interlocked pair of flux rings (see, e.g.,
the first panel of Fig. \ref{fig:bbiso}), 
for which the magnetic helicity is given by twice the product of the two
magnetic fluxes of each of the two flux rings. However, helicity
is also associated with two orthogonal flux tubes as shown
in \Fig{fig:ortho}.

The dramatic difference between the dissipation
of magnetic energy and magnetic helicity can best be seen by
contrasting the equations of the conservation of magnetic energy
and magnetic helicity,
\EQ
\half{\dd\over\dd t}\bra{\BB\cdot\BB}
=-\bra{\uu\cdot(\JJ\times\BB)}-\eta\bra{\JJ\cdot\JJ},
\label{eq:ddBB}
\EN
\EQ
\half{\dd\over\dd t}\bra{\AAA\cdot\BB}
=-\bra{\uu\cdot(\BB\times\BB)}-\eta\bra{\JJ\cdot\BB},
\label{eq:ddAB}
\EN
where angular brackets denote volume averages, and surface terms are
assumed to vanish. 

The important point to note here is that the magnetic energy
can maintain a steady
state where Joule dissipation, $\eta\bra{\JJ^2}$, can be
finite and large if work is done against the Lorentz force, i.e.\ if
$-\bra{\uu\cdot(\JJ\times\BB)}>0$. At the same time, however, there is
no such term in the magnetic helicity equation, so in that case a steady
state is only possible if the current helicity, $\bra{\JJ\cdot\BB}$,
vanishes.

In the absence of any forcing, $\ff$, the hydrodynamic helicity is conserved
in a similar manner, but there are two important differences. First,
because $\int\uu\cdot\oo\,\dd V$ contains one more derivative
than $\int\uu^2\,\dd V$, it dissipates faster than the energy
if there is dissipation. Second, if there is forcing, the hydrodynamic
helicity is no longer conserved. This difference to the hydromagnetic
case can best be seen by contrasting \Eqs{eq:ddBB}{eq:ddAB} with the
corresponding equations in hydrodynamics,
\EQ
\half{\dd\over\dd t}\bra{\uu\cdot\uu}
=\bra{\uu\cdot\ff}-\nu\bra{\oo\cdot\oo},
\label{eq:dduu}
\EN
\EQ
\half{\dd\over\dd t}\bra{\uu\cdot\oo}
=\bra{\oo\cdot\ff}-\nu\bra{\ttheta\cdot\oo},
\label{eq:dduo}
\EN
where $\ttheta=\nab\times\oo$ is the curl of the vorticity, and surface
terms are again assumed to vanish. Thus, unlike the
magnetic counterpart, kinetic helicity conservation is only possible in
the special case where the forcing is perpendicular to the vorticity
and the flow is inviscid. With dissipation and 
the absence of forcing both kinetic energy and kinetic helicity are
decaying, but kinetic helicity contains an extra derivative more
than the kinetic energy and so decays faster than energy and does not
pose a hard constraint.

\section{Interlocked flux rings}

\begin{figure}[t]
\begin{center}
\includegraphics[width=.99\textwidth]{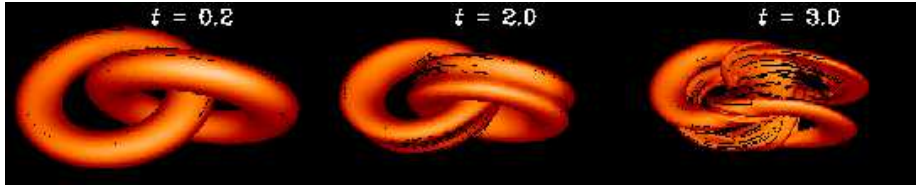}
\end{center}
\caption[]
{Resistive evolution of an initially interlocked pair of flux rings.
Isosurfaces of the magnetic field are shown at different times.}
\label{fig:bbiso}
\end{figure}

In \Fig{fig:bbiso} we show an example of an initial
flux tube configuration where, in our case, each tube has the flux
$\Phi=\int\BB\cdot\dd\SSS=0.7B_0d^2$, where $B_0$ is the maximum field
strength in the core of each tube and $d$ is its radius. The magnetic
helicity is measured to be $H=\int\AAA\cdot\BB\,\dd V=0.98B_0^2d^4$,
in perfect agreement with the formula $H=2\Phi^2$.

\begin{figure}[t]
\begin{center}
\includegraphics[width=.99\textwidth]{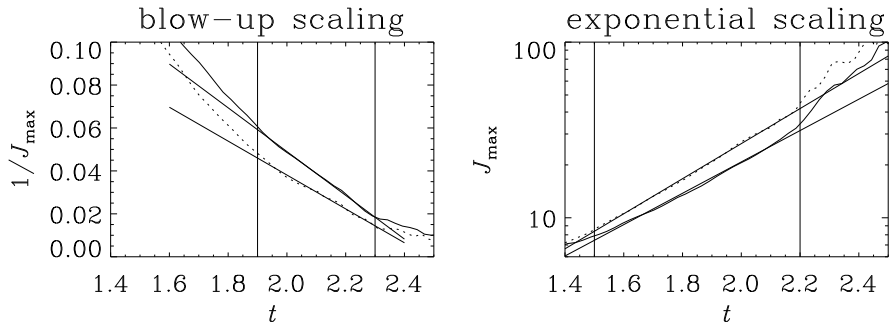}
\end{center}
\caption[]
{Semi-logarithmic plot of $\|\JJ\|_\infty$ for a compressible $240^3$
calculation in a domain of size 4 (dotted line: filtered, and solid
line: unfiltered initial conditions) together with fits to exponential
growth and blow-up behavior, respectively. The blow-up scaling fits better
at later times.}
\label{fig:pjmax}
\end{figure}

\begin{figure}[t]
\begin{center}
\includegraphics[width=.63\textwidth]{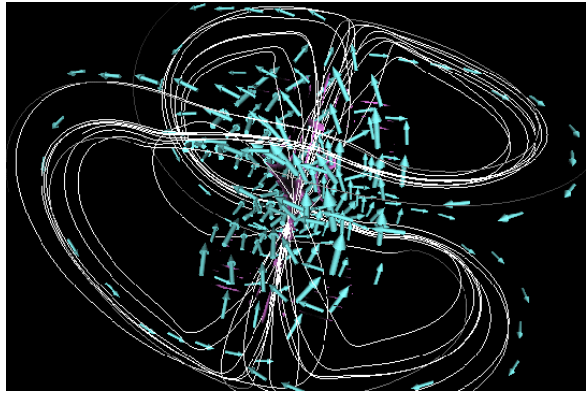}
\end{center}
\caption[]{Magnetic field lines (white) together with some field
vectors (in grey) indicating the field orientation for a resistive
flux ring calculation at $t=4$, i.e.\ shortly after the time of
the suspected singularity in the ideal calculation.}
\label{fig:ssnap1}
\end{figure}

The subsequent evolution of this flux tube configuration is governed
by the curvature force acting separately in each flux tube trying to
make them contract. Eventually the two tubes come into contact
and produce an intense current sheet where they touch.
Figure \ref{fig:pjmax} shows the peak current, $\|\JJ\|_\infty$, plotted in
two ways, one showing a period of singular growth and the
other showing a period of exponential growth.
This initial period is represented in Fig. \ref{fig:bbiso} by
the first two frames showing the approach and initial
deformation of the linked flux tubes.
At the last time visualized, the surfaces in the outer region
appear to merge into a single continuous flux tube, while the
inner region appears to be annihilated in a complicated reconnected
structure with writhe.
Figure \ref{fig:ssnap1} takes another look at this time using flux
lines instead of surfaces.  The flux lines in the
outer region that appeared
to be continuous can now be seen to change direction abruptly where they plunge
into the inner region.  And the inner region is now seen to be continuously
connected to the outer flux lines and instead of being a single flux
tubes with writhe, it now appears to be the original flux lines just
twisted around each other with almost no reconnection.

In the ideal case,
the magnetic helicity is conserved for all time. Even in the resistive
case the magnetic helicity is very nearly constant.  Furthermore, the
peak current increases to large values, which appear to be limited only
by the numerical resolution.
This is related to the newly posed millennium question of whether
regularity of the Navier-Stokes equations can be shown \cite{claymath}.
A singularity probably does not develop for the full viscous and
resistive equations due to the development of reconnection.
However, singularities do seem possible for Euler and ideal MHD.
Numerical calculations have been used to provide insight into the
interaction of anti-parallel vortex tubes using the incompressible
Euler equations \cite{Kerr93}.  The key to providing useful results
was the direct comparison with hard analytic bounds for the maximum
growth rate of the vorticity \cite{BKM84}.  
This initial condition was very contrived with special symmetries
and no helicity, unlike real flows, and its generality
remains uncertain.

A similar analytic bound has been shown to exist for the ideal
MHD equations \cite{CKS97}.  Therefore two intertwined
questions have arisen.  First, is there an initial condition for
ideal MHD which might show similar singular growth? Second, what role
might the various types of helicity play in suppressing or enhancing
the reconnection rate?  It has been found \cite{KerrB99} that
in the ideal case the
two interlocked magnetic flux tubes go through a phase where
behavior consistent with a singularity seems plausible.  This
was surprising because as noted above this is a nearly maximally
helical initial condition and was expected to suppress nonlinearity.
For hydrodynamics, it has also been claimed \cite{BoratavPZ92}
that the helical initial 
condition of two orthogonal vortex tubes showed signs of a singularity,
although the analytic test \cite{BKM84} was never applied.  These
two cases raise two possibilities. In hydrodynamics there might exist
a mechanism whereby helicity is shed permitting stronger nonlinear growth,
while in MHD the nonlinearity really comes from 
${\bf J}\times {\bf B}$ and ${\bf J}\cdot {\bf B}$ 
is not conserved, so there might in fact be no
constraint upon locally strong nonlinearity.

Let us consider an argument for why helicity might
be required to allow, at least for a period, nearly singular growth
for ideal MHD.  This is based upon old arguments for why
there could not be a singularity of Euler.
It has been argued that a singularity would not
occur for Euler because what drives the growth in the
vorticity is the axial strain stretching the vorticity, and this
strain must grow at the same rate as the vorticity to sustain this
growth.  This could only be achieved by an enormous growth in the
curvature of vortex lines \cite{GibbonHeritage97}, which in turn
would require a delicate balance in the growth of the pressure Hessian.
This was originally thought not to be feasible, but newer analysis of
the anti-parallel Euler calculations \cite{Kerr96,Kerr98}
has shown that all of the analytic
requirements needed to achieve this delicate state are in fact obeyed.
This is possible because the vorticity and the strain are in fact
just different manifestations of the same vector field and
can be strongly aligned.

While it has not been shown analytically, one would expect
that for ideal MHD to show similar singular growth,
a similar delicate balance would have to exist between the vorticity,
the current, and the magnetic and velocity strain fields.  Current
and magnetic strains are just different manifestations of the
same vector magnetic field, just as vorticity and strain are manifestations
of the same vector velocity field.  However, one is still left
with the current being completely distinct from vorticity.  Only
if there is some property of the vector fields that strongly couples these
two fields could they act in concert to give a singularity.
Perhaps because helicity is conserved, for strongly helical structures
there exists such a constraint.  Therefore only for strongly
helical magnetic structures could singular, or nearly singular,
nonlinear growth occur.

New analysis has shown that the location of the peak in the
current is at the juncture between the outer flux lines and the inner flux lines
where the maximum in the curvature is located.  This would be consistent
with new mathematical analysis by J. D. Gibbon (unpublished)
that strong magnetic field
line curvature should be associated with any singular growth.
The vorticity is also the strongest in this region, suggesting the
type of symbiotic growth of current and vorticity that we believe
is needed if there is to be a singularity of ideal MHD.

Figure \ref{fig:ssnap1} is resistive, but is very similar to visualizations
of new ideal calculations that were run at higher resolution, up to the
equivalent of $1296^3$ mesh points if a uniform mesh had been used.
However, these new calculations, while they do extend the period of
seemingly singular growth, now appear to show that the singular growth
eventually is suppressed in the incompressible case.  
The evidence relies on consistent behavior
between the two highest resolution calculations.  

What might be the cause of this suppression?  While it will take time
to fully understand these massive data sets, the initial indications
are that it is occurring as the peak in the current moves outside
the inner region with its strongly aligned vorticity, magnetic, and
current fields.  Our suspicion is that the importance of the inner
region is that, through twist, the location of most of the initial
helicity associated with the linked flux tubes is in the inner region.
If this can be shown, then it might tell us that the secret
to maintaining nearly singular growth is to maintain as high a level of local
helicity for as long as possible.  This would be consistent with 
arguments \cite{Parker88} that fast reconnection is associated with
the entanglement of flux lines due to footpoint motion, which is known
to produce the required heating rates \cite{GalsNord96}.
Helicity has also
been shown to play a role in coronal simulations of an arcade
and a twisted flux loop \cite{Amari00}, 
with nearly singular growth in current
similar to what we have observed.  

There are other possible mechanisms that could suppress singular growth.
In a simulation \cite{GrauerMar00} of nearly the same initial condition
as the one used in our original paper \cite{KerrB99}, there is only exponential
growth that is associated with the appearance of current
sheets.  More recent detailed analysis of our calculation shows that
the exponential growth is actually associated 
with the appearance of two nearly overlapping
orthogonal current sheets and the pressure barrier between
them that suppresses stretching terms and growth.  This is an important
result because in some sense the more physical initial situation might be 
two flux tubes that do not overlap at all.  Our simulations
that show stronger growth in the current 
all have some overlapping between the
initial flux tubes, something that should not happen in an astrophysical
situation where the flux tubes are initially separated by large distances.

\section{Orthogonal vortex tubes}

\begin{figure}[t!]\begin{center}
\includegraphics[width=.55\textwidth]{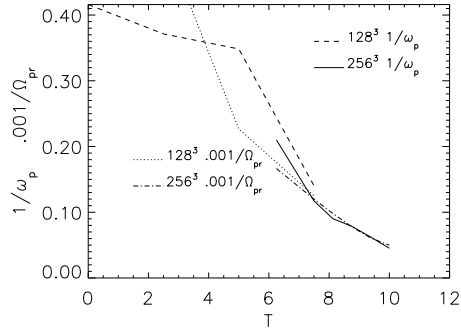}
\end{center}
\caption[]{$1/\|\omega\|_\infty$ and
$1/\langle\omega_i e_{ij}\omega_j\rangle$
in Euler for orthogonal vortices.}
\label{fig:sk2om}
\end{figure}

\begin{figure}[t]\begin{center}
\includegraphics[width=.99\textwidth]{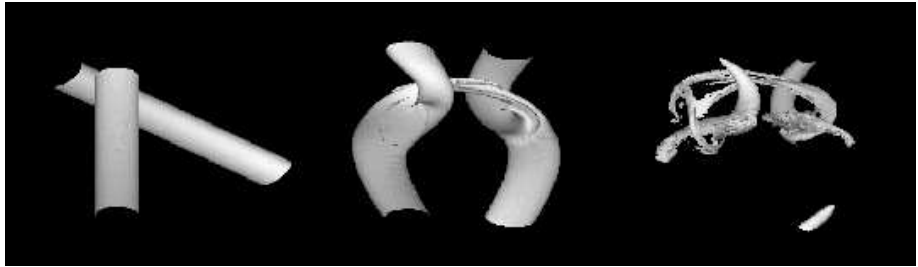}
\end{center}
\caption[]{Isosurfaces of vorticity as a fraction of the peak vorticity.
The three frames are
$t=0$, 6 and 10.  Arms are pulled out of the original vortices,
become anti-parallel, then vorticity within the arms develops singular
behavior.}
\label{fig:ortho}
\end{figure}

We now turn to the case of {\it straight} tubes that are orthogonal
to each other. We note that also in this case there is finite helicity.
The magnetic case has been studied previously \cite{Dahlburg95},
but here we focus on the hydrodynamic case with vortex tubes.
Figure \ref{fig:sk2om} shows the inverse of the peak vorticity
and the inverse of the enstrophy production rate for the orthogonal
vortex tubes whose inviscid evolution is shown in Fig. \ref{fig:ortho}.
Plotting these inverses was previously shown to be the most effective
way to highlight the $1/(t_c-t)$ singular behavior.
Figure \ref{fig:sk2om} shows that the initial growth is weak,
unlike the anti-parallel case.  Then the
growth of peak vorticity, $\omega_{\rm p}$, and enstrophy production,
$\Omega_{\rm pr}$, becomes stronger
with their inverses going roughly linearly to zero at the same
singular time.

What is the configuration around the peak vorticity once singular
growth starts?  And what role does helicity play?  Analysis of the
three-dimensional fields shows that the peak vorticity is located
in the arms that are being pulled off of the two original orthogonal
vortices.  The last time shows that these isosurfaces are parallel,
and analysis shows that the vorticity within these surfaces is 
anti-parallel.  That is, to develop singular growth exactly the
same alignment of vorticity that was previously described as a
contrived situation is actually what the dynamics generate by
themselves.  This is consistent with vortex filament work
\cite{PumirSiggia87}.  In terms of helicity, locally around the
anti-parallel vortices there is no kinetic helicity density. Therefore in
order for orthogonal vortices to develop singular growth, the flow
must realign itself to be non-helical, shedding any helicity to
achieve this.

In conclusion, these calculations have demonstrated that the role of
helicity can be rather complex.  In the hydrodynamic
case it appears that the absence of helicity is required for there
to be singular growth and in the MHD case helicity seems to be required.
The role of helicity upon reconnection should now be investigated
for these and similar configurations \cite{Dahlburg95}.

%

\end{document}